\documentclass[aps,prl,preprint,groupedaddress]{revtex4-1}

\usepackage{graphicx,graphics}
\usepackage{caption}
\usepackage{subcaption}

\begin{document}


\title{Ranking the Economic Importance of Countries and Industries}



\author{Wei Li} \affiliation{Center for Polymer Studies and Department of Physics,
    Boston University, Boston, USA}
\author{ Dror Y. Kenett} \affiliation{Center for Polymer Studies and Department of Physics,
    Boston University, Boston, USA}
 \author{Kazuko Yamasaki}\affiliation{Department of Environmental Sciences, Tokyo
   University of Information Sciences, Japan}
\author{H. Eugene Stanley}\affiliation{Center for Polymer Studies and Department of Physics,
    Boston University, Boston, USA}
 \author{Shlomo Havlin}\affiliation{Department of Physics, Bar-Ilan University, Israel}


\date{\today}

\begin{abstract}
In the current era of worldwide stock market interdependencies, the
global financial village has become increasingly vulnerable to systemic
collapse. The recent global financial crisis has highlighted the
necessity of understanding and quantifying interdependencies among the
world's economies, developing new effective approaches to risk
evaluation, and providing mitigating solutions. We present a
methodological framework for quantifying interdependencies in the global
market and for evaluating risk levels in the world-wide financial
network. The resulting information will enable policy and decision
makers to better measure, understand, and maintain financial
stability. We use the methodology to rank the economic importance of
each industry and country according to the global damage that would result from
their failure. Our quantitative results shed new light on China's
increasing economic dominance over other economies, including that of
the USA, to the global economy.\end{abstract}


\maketitle

\section{Introduction}

The growth of technology, globalization, and urbanization has
caused world-wide human social and economic activities to become
increasingly interdependent
\cite{helbing2010fundamental,havlin2012challenges,san2012challenges,helbing2012systemic,lazer2009life,king2011ensuring,lorenz2011social,yamasaki2012complex,meng2009production,rinaldi2001identifying,solomon2003pioneers,levy2010scale,klimek2012empirical}.
From the recent financial crisis it is clear that components of this
complex system have become increasingly susceptible to collapse.
Integrated models, currently in use, have been unable to predict instability,
provide scenarios for future stability, or control or even mitigate
systemic failure. Thus, there is a need of new ways of quantifying complex system
vulnerabilities as well as new strategies for mitigating systemic damage
and increasing system resiliency
\cite{farmer2009economy,lux2009economics}. Achieving this would also
provide new insight into such key issues as financial contagion
\cite{forbes2001measuring,forbes2002no} and systemic risk
\cite{bodie2002investments,billio2010econometric,bisias2012survey} and would provide a way of
maintaining economic and financial stability in the future.

It is clear from the recent crisis that different sectors of the economy
are strongly interdependent. The housing bubble in the USA caused a
liquidity freeze in the international banking system, which in turn triggered a
massive slowdown of the real economy costing many trillions of dollars
and threatening the financial integrity of the European Union. This
demonstrates the high level of dependence between different components in
the world economic system. Because strong nonlinearities and feedback
loops make economic systems highly vulnerable, we need to understand the
behavior of the interacting networks that comprise the economy. How do
they interact with each other, and what are their vulnerabilities?
One may consider an industry in two countries respectively, for example,
electrical equipment industry in China and the USA, and ask which industry is more important to economic stability, electrical equipment in China or
 that in the USA? Is electrical equipment in China
 more critical for global economy stability when the production in
 electrical equipment in China is reduced, and how it will impact the industries
 within the country and industries abroad.

To answer these questions, we employ recent advances in the theory of cascading
failures in interdependent networks
\cite{buldyrev2010catastrophic,gao2011networks,li2012cascading}. In the
case of interdependent networks, a malfunction of only a few components
can lead to cascading failures and to a {\it sudden collapse\/} of the
entire system. This is in contrast to single isolated networks, which
tend to {\it collapse gradually\/}
\cite{parshani2010interdependent}. These recent results indicate the
central importance of interconnectivity and interdependency to the
stability of the entire system. There
have been studies of the complex set of coupled economic networks
\cite{garas2010worldwide,schweitzer2009economic,huang2013cascading,hidalgo2008network}.
However, the importance of countries and
industries in the stability of the global economy have not been analyzed, and there is a need for
useful methods to quantify and rank their economic importance and influence.

The input-output (IO) model is a technique that quantifies
interdependency in interconnected economic systems. Wassily Leontief
\cite{leontief2004wassily} first introduced the IO model in 1951, for which he received the Nobel Prize in Economics in 1973. It can be used to study
the effect of consumption shocks on interdependent economic systems
\cite{isard1960methods,lahr2001input,ten2005economics,miller2009input,santos2006inoperability,pokrovskii2012econodynamics,leontiev1986input}. Analysis
of IO data is performed using such techniques as the hypothetical
extraction method (HEM)
\cite{miller2001taxonomy,temurshoev2010identifying}.  Although HEM can
measure the relative stimulative importance of a given industry by
calculating output with and without the industry being examined, it does
not quantify each industry's full spectrum of importance to the stability of global
economic system. For example, if an industry in a given country
collapses completely due to natural disaster or civil unrest it will no
longer be able to consume products supplied by other industries.  This
can cause cascading failure in the economic system if the other
industries cannot function when the cash flow from the failed industry
is removed. By measuring how widely the damage spreads, we will rank here an industry's importance within the world-wide economic system.

In this paper we examine the interdependent nature of economies between and within 14 countries and the rest of the world (ROW). We use an
``input-output table'' \cite{timmer2012world} and focus on economic
activity during the period 1995--2011. From the table we analyze data from 14
countries (Australia, Brazil, Canada, China, Germany, Spain, France, UK,
India, Italy, Japan, Korea, Russia, and USA) and from the rest of the
world (ROW). The economic activity in each country is divided into 35 industrial classifications. Each cell in the table shows the output composition of
each industry to all other 525 industries and its final demand and export to
the rest of the world (see Ref.~\cite{wiotdata}). 
We construct an output
network using the 525 industries as nodes and the output product values
as weighted links based on the input-output table, and focus on the
output product value for each industry.

Our goal is to introduce a methodology for quantifying the importance of
a given industry in a given country to global economic stability with respect to
other industries in countries that are related to this industry.
Thus, we study the inflow and
outflow of money between each set of 35 industries and the ROW in each of
the 14 countries (see the Data section for more detail). We use the
theory of cascading failures in interdependent networks to gain valuable
information on the local and global influence on global stability of different economic
industries, a methodology that can provide valuable new insights and
information to present-day policy and decision makers.

\section{World Input-Output Table Data}

The database we use in this study is the World Input-Output Table (WIOT)
\cite{wiotdata}. It provides data for 27 European countries, 13 other
countries, and the rest of the world (ROW) for the period
1995--2011. For our sample, we select the 14 countries with the largest
domestic input and the largest import in production in 2011 and also the
ROW (adding the remaining 26 countries into the ROW). Using this sample,
we construct a new input-output industry-by-industry table. For sake of
simplicity, we assume that each industry produces only one unique
product. In the WIOT, the column entries represent an industry's inputs
and the row entries represent an industry's outputs. The rows in the
upper sections indicate the intermediate or final use of products. A
product is intermediate when it is used in the production of other
products (intermediate use). The final use category includes domestic
use (private or government consumption and investment) and exports. The
final element in each row indicates the total use of each product. The
industry columns in the WIOT contain information on the supply of each
product. The columns indicate the values of all intermediate, labor, and
capital inputs used in production. Total supply of the product in the
economy is determined by domestic input plus final demand. The IO table contains negative
numbers as outputs for various reasons \cite{timmer2012world}, but their fraction is fairly small and their values are small as well.
For simplicity, we set these negative numbers as zero in our analysis.

Based on the supply of the product for each industry, it is possible to construct a
directed product supply network. Then we reverse the direction of the links in the
network (see schematic representation in
Fig.~1), and the network represents the money
outflow from one industry to
another; e.g., electrical equipment industry is pointing to machinery
industry because electrical equipment industry
buys a product from machinery industry. The links are weighted according to
 the value of products from
machinery industry as its output electrical equipment industry as its input.

\section{Industry tolerance}

In order to quantify and rank the influence of industries in the stability of this
global network, we perform a cascading failure tolerance analysis
\cite{buldyrev2010catastrophic}. Our model is described as follows.
Suppose industry A fails, other
industries can no longer sell their products to industry A and thus they
lose that revenue. The revenue industry B is reduced by a
fraction $p^\prime$, which is defined as industry B's revenue
reduction caused by the failure of industry A divided by that industry B's
total revenue. The tolerance fraction $p$ is the threshold above which
an industry fails. This occurs when reduced revenue fraction $p^\prime$
is larger than tolerance fraction $p$. Here we assume that (i) $p$ is
the same for all industries and that every industry fails when its
$p^\prime>p$ and (ii) the failure of an industry in country A does not
reduce the revenue of the other industries in the same country A
because they are able to quickly adjust to the change.

The methodology can be schematically illustrated as follows: In step 1, industry $A$ in country $i$ fails. This causes other industries in
other countries to fail if their $p^\prime > p$.
Assume that in step 2 industries $B$, $C$, and
$D$ fail. The failure of these industries in step 2 will reduce
 other industries' revenue and cause more industries including those in
 country $i$ to have a reduced fraction $p^\prime$.
 Thus in step 3 there is an increased number of industries whose
 $p^\prime > p$. Eventually the system reaches a steady state in
which no more industries fail. The surviving industries will all have
a reduced revenue fraction that is smaller than the tolerance fraction,
i.e., $p^\prime\leq p$. Figure~1 demonstrates this process,
and the subfigures show the steps in the cascading failure.

\begin{figure}
\centerline{\includegraphics[width=0.5\textwidth]{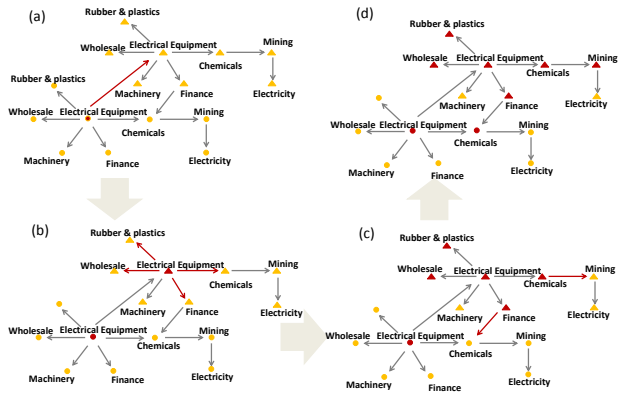}}
\caption{Schematic representation of each step in the cascading failure propagation in
  the world economic network. We present an example of two countries,
  where circle nodes represent country 1, and triangles represent
  country 2. Both countries have the same industries, and the arrow
  between two nodes points in the direction of money flow.  (a) There is
  a failure in electrical equipment
  industry in Country 1 (circle) which causes a failure of electrical
  equipment industry in Country 2. In (b) the failure of electrical
  equipment industry in Country 2 will cause rubber and
  plastics, wholesale, finance and chemicals in Country 2 to fail.
  (c) Shows mining in Country 2 and chemicals in Country 1 to fail further.
  (d) The network reached a steady state. The red nodes
  represent the failing industries and the yellow nodes stand for the
  surviving industries.}
\label{IOpic24.eps}
\end{figure}

To determine how much the failure of each industry would impact the
stability of the economic network, we change the tolerance fraction $p$
from 0 to 1 and measure the fraction of surviving industries left in the
network. When the tolerance fraction approaches 0, any revenue reduction
caused by the failure of one industry can easily destroy almost all the other
industries in the network, and the network will collapse. When the
tolerance fraction approaches 1, all the industries can sustain a large
reduction of revenue and the failure of one industry will not affect the
others.

Figure ~2(a) shows the failures of electrical equipment industry in China
and energy industry in the United States for the 2009 WIOT
and shows the fraction of the largest cluster of
connected and functional industries as a function of the tolerance fraction $p$ after
the Chinese electrical equipment industry becomes malfunction and is removed from the network
due to a large shock to this industry.
The shock could result from different causes, such as natural environmental disaster,
 government policy changes, insufficient financial capability. The removal of China
 electrical equipment industry will cause revenue reduction in other industries because China
 electrical equipment industry is not able to buy products and provide money to other industries.
 When $p$ is small, the
industries are fragile and sensitive to the revenue reduction, causing most of
the industries fail, and the number of the surviving industries is very
small. When
$p$ is large, the industries can tolerate large revenue reduction and
are more robust when revenue decreases. The number of the surviving industries
tends to increase abruptly at a certain $p=p_c$ value as $p$ increases. Figure~2(b)
shows the number of cascading steps that elapse before a stable state is reached
as a function of tolerance fraction $p$ after removing the Chinese electrical
equipment industry or the USA energy industry.
The number of steps reaches a peak when $p$
approaches criticality $p_c$\cite{parshani2010interdependent}.

\begin{figure}
\centering
\includegraphics[width=\textwidth]{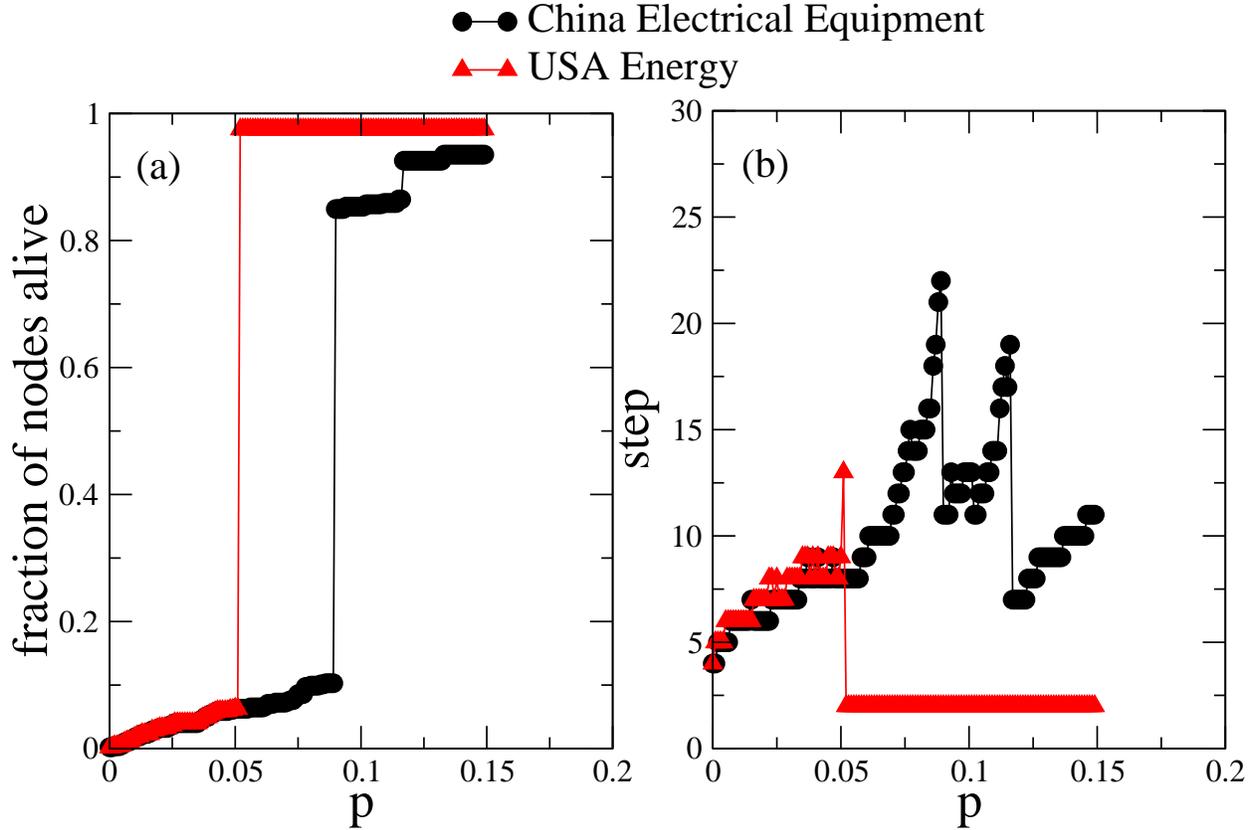}
\caption{Typical examples of industry tolerance threshold $p_c$.
 Tolerance threshold
  $p_c$ shows the importance of each industry in the international
  industry networks. In (a) the black curve shows the fraction of
  surviving industries as a
  function of tolerance threshold for the case when the electrical equipment
  industry in China fails in year 2009 and the red curve represents the case of the failure of
  energy industry in 2009 in the USA. (b) Shows the number of failure steps as a
  function of $p$ corresponding to (a). The total number of steps is the number of
  cascades it
  takes for the network to reach a steady state after certain initial
  failure. The peaks in (b) are corresponding to the abrupt jumps in (a),
   which means that large numbers of cascade steps are
   associated at $p_c$ with a dramatic failures in the industry network.
   Each step in the cascading process in (b)
   is demonstrated in
  Figure~1. }
\label{chinaChemical}
\end{figure}

Here we analyze for each of the 525 industries this important parameter, the critical tolerance
threshold $p_c$. Our goal is to determine the threshold $p_c$ at which global
network subject to collapse becomes stable and most of
the industries in the network survive after initial failures. To this end,
we assume that the $p_c$ is the critical threshold below
which less than 30\% of industries survive.
When $p>p_c$, more than 30\% of the remaining
industries survive after the failure cascade in the system is over. When
$p<p_c$, the survival rate of the remaining industries is 30\% or less.
The higher this threshold,
the higher the impact of a failing industry will be on the vulnerability
of the global network. Without losing generalization, we
also take different fraction of surviving industries to set $p_c$ in simulations. For example, when the $p_c$ is the threshold below
which 50\% of industries survive, the importance of industries and countries maps are shown as in supplementary information.
The correlation of
$p_c$ values in simulations using 50\% fraction of surviving industries and 30\% surviving industries is 1 and the $p_c$ are identical in
both scenarios. Thus, we find that the critical threshold $p_c$ is not sensitive to the chosen fraction in this industry network, for
values below 50\%. The following $p_c$ in this study is defined as the critical threshold below
which less than 30\% of industries survive.

We use this methodology to test how the failure of an
individual industry in a given country impacts the stability of the entire system. Thus, $p_c$ of an
industry is our measure of the importance of this industry in the
global economic network.

Using the tolerance threshold $p_c$, we can quantify and rank the
economic importance of each industry. We measure the tolerance threshold
of 35 industries in 14 countries between 1995 and 2011. We calculate
the tolerance of each industry according to how much it affects the
entire network, i.e., all 35 industries in all 14 countries.

\section{Importance of Country and Industry}

The proposed methodology provides the means to quantify and rank the importance
of each industry to the stability of the global economic network or
the importance of each country in the
global economic network.

We define the importance of each country
by averaging all the 35 industries $p_c$ within the country
for a specific year,

\begin{equation}
I_{country}(i)\equiv \frac{1}{n}\sum_{k=1}^n{p_c(i,k)}
\label{eqctry}
\end{equation}
where $n$ is the number of industries. To rank the importance of a given
country, we average the tolerance across industries as shown
in Eq.~\ref{eqctry}.

Figure~3(a)
shows the importance of all countries for different years where
the average in Eq.~\ref{eqctry} is taken over the four largest $p_c$ values,
in order to consider the strongest industries. Note that
the rest of the world (RoW) is the most
important ''country'' since it includes many countries other
than the 14 countries included in the sample. Countries in RoW provide
products that are crucial inputs to these 14 countries
and their industries. We also define the importance of the individual industries in
terms of the average of their $p_c$,

\begin{equation}
I_{industry}(i)\equiv \frac{1}{T}\sum_{j=1}^T{p_c(i,t)},
\label{eqindustry}
\end{equation}
where $T$ denotes years. Figure~3(b) shows the
average tolerance of industry over all years (1995-2011)
 for the individual countries. For the sake of
clarity, we plot only the top 20 industries with respect to the average of
$I_{\rm industry}$ values for all the countries.
Note that the electrical equipment industry is the
most important when we average the tolerance fraction $p_c$ across 17
years. The energy industry $p_c$ in the United States, for example, is relatively
high because the United States is the world's largest energy consumer.

\begin{figure}
\centering
\begin{subfigure}{0.4\textwidth}
\centering
\includegraphics[width=0.8\textwidth]{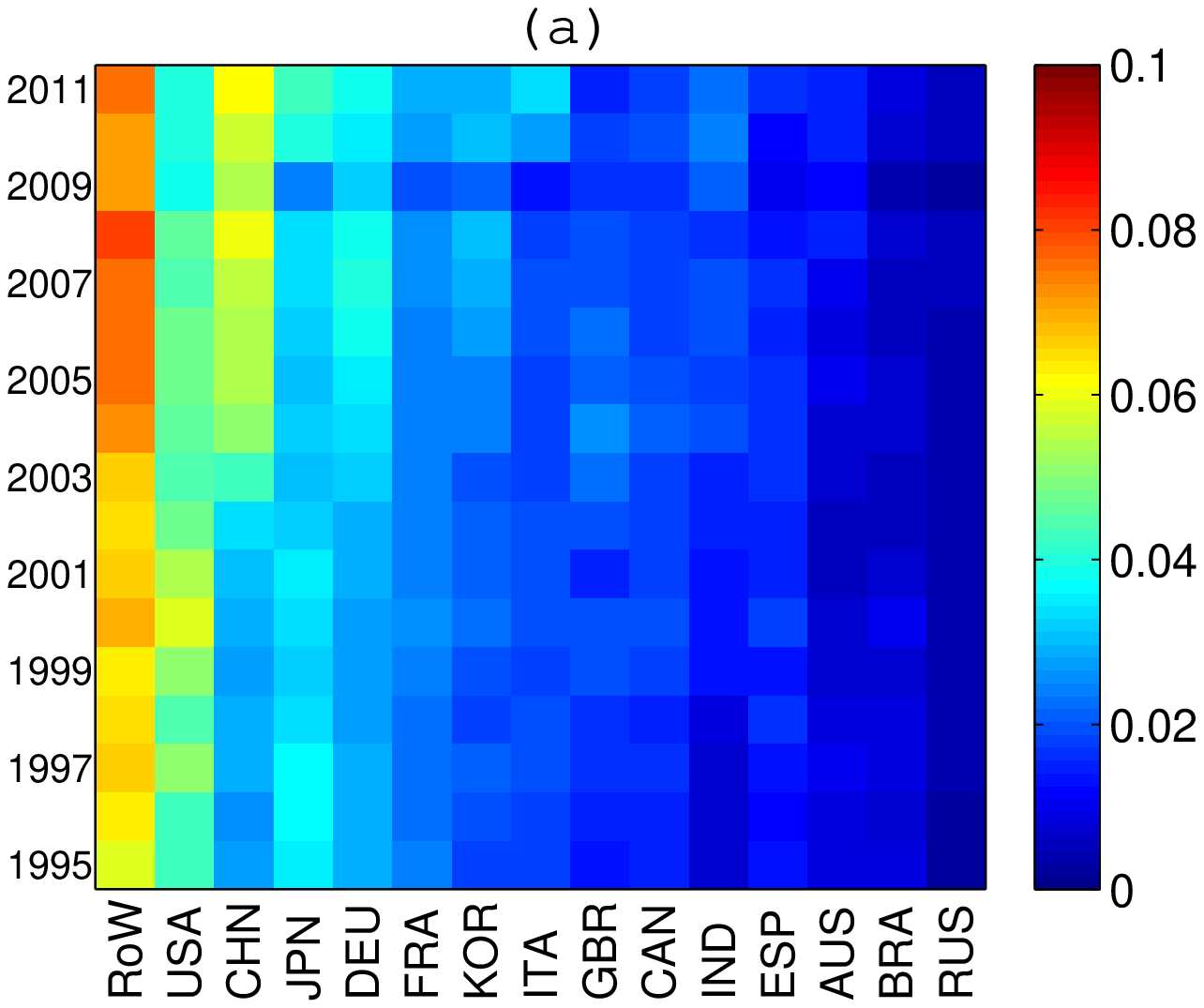}
\label{fig:year}
\end{subfigure}
\begin{subfigure}{0.4\textwidth}
\centering
\includegraphics[width=0.8\textwidth]{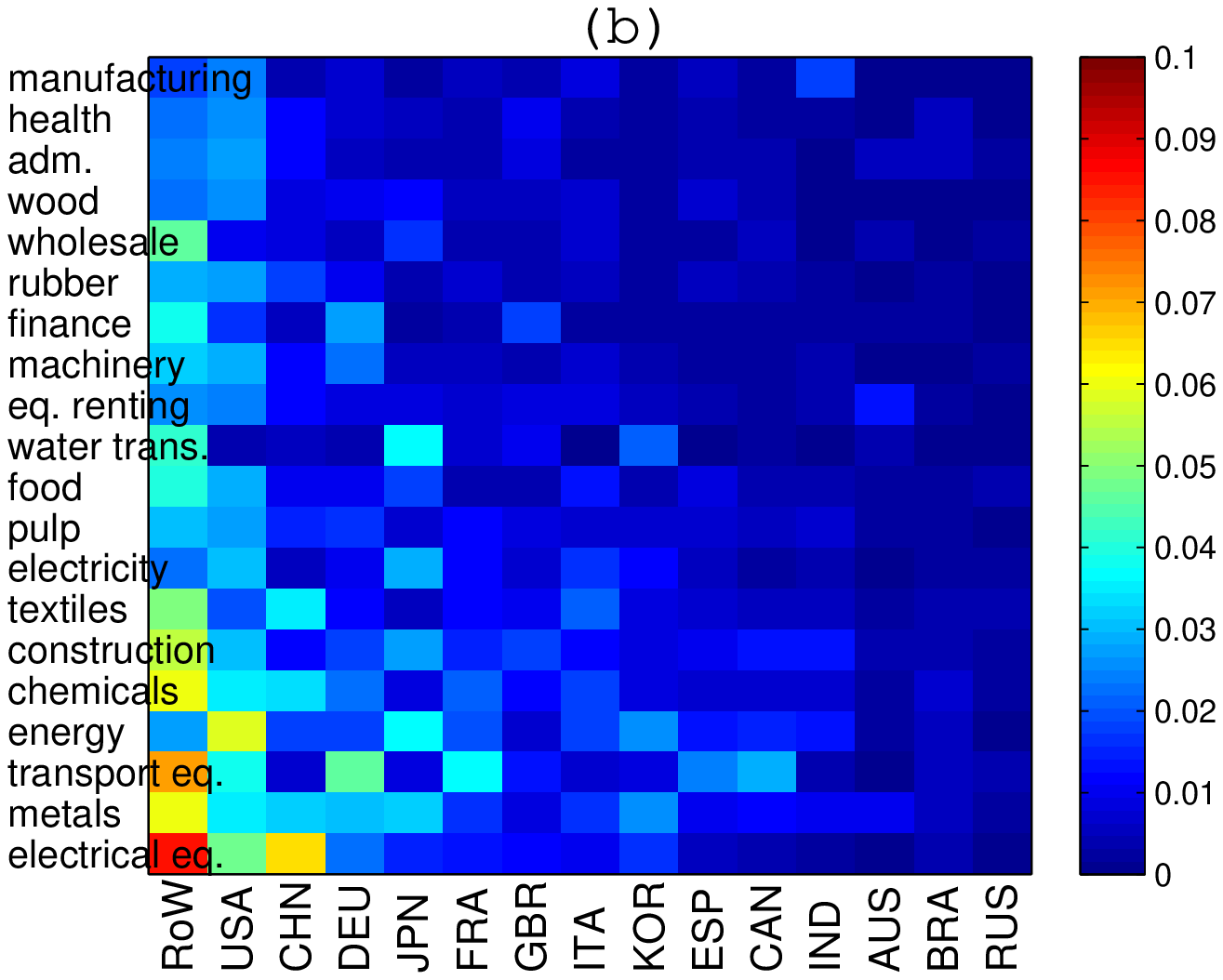}
\label{fig:ind}
\end{subfigure}
\caption{Ranking the importance of individual industries or
  countries. (a) Map of averaging industry
  tolerances for each country,
  $I_{country}$. The average tolerance $p_c$ is calculated by averaging
  the largest four $p_c$ for each country in one year. The color
  represents the strength of the tolerance, ranging from blue for low
  tolerance, to red for high tolerance. (b) Averaging industry tolerance
  over 17 years (1995-2011). The average tolerance $p_c$ is calculated
  by averaging the industry tolerance over 17 years for each country
  (for presentation sake, we plot only the top 20 industries with
  respect to their $I_{industry}$ value). The values are represented
  using the same color code.}
\label{ctrymapIndmap
}
\end{figure}

\section{Robustness and stability of economic structure}

We use the critical tolerance threshold $p_c$ to measure the importance
of each industry in the global economic network. We can also rank each
industry according to their tolerance $p_c(i,k)$ for each separate
country---the tolerance of industry $i$ in country $k$. By comparing
industry order rankings for different years, we can study the similarity
in economic environment across a period of 17 years.

To do this we use the Kendall $\tau$ coefficient \cite{kendall1938new},
which measures rank correlation, i.e., the similarity in data orderings
when ranked by each quantity. Let $(x_1, y_1), (x_2, y_2), \dots, (x_n,
y_n)$ be a set of observations of the random variables $X$ and $Y$
respectively. Any pair of observations $(x_i)$ and $(y_i)$ is concordant
if the ranks for both elements agree, that is if $x_i<x_j$ and $y_i<
y_j$ or if $x_i>x_j$ and $y_i>y_j$. Otherwise the pair is discordant. If
$x_i=x_j$ or $y_i=y_j$, the pair is neither concordant nor
discordant. The Kendall $
\tau$ coefficient is defined as

\begin{equation}
\tau = \frac{n_{+} - n_{-}}{\frac{1}{2} n (n-1) },
\label{kendall.eq}
\end{equation}

\noindent where $n_{+}$ is the number of concordant pairs and $n_{-}$ is the
number of discordant pairs. The coefficient $\tau$ is in the range
$-1\le\tau\le1$. When the agreement between the two rankings is perfect,
the coefficient is 1. When the disagreement between the two rankings is
perfect, the coefficient is $-1$.  If $X$ and $Y$ are independent, the
coefficient is approximately zero.

\begin{figure*}
\begin{subfigure}{0.3\textwidth}
\centering
\includegraphics[width=0.7\textwidth]{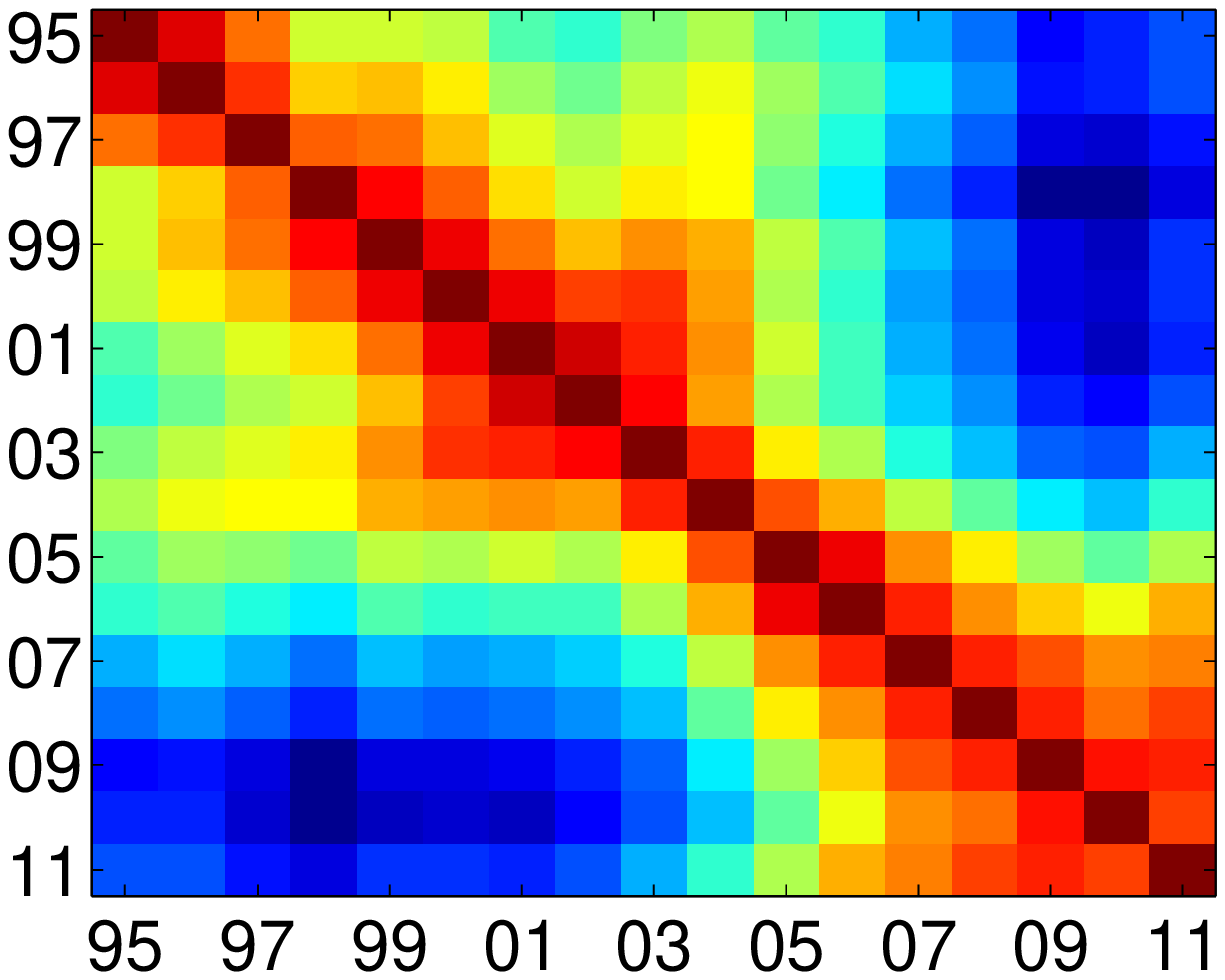}
\caption{China}
\label{fig:china}
\end{subfigure}
\begin{subfigure}{0.3\textwidth}
\centering
\includegraphics[width=0.7\textwidth]{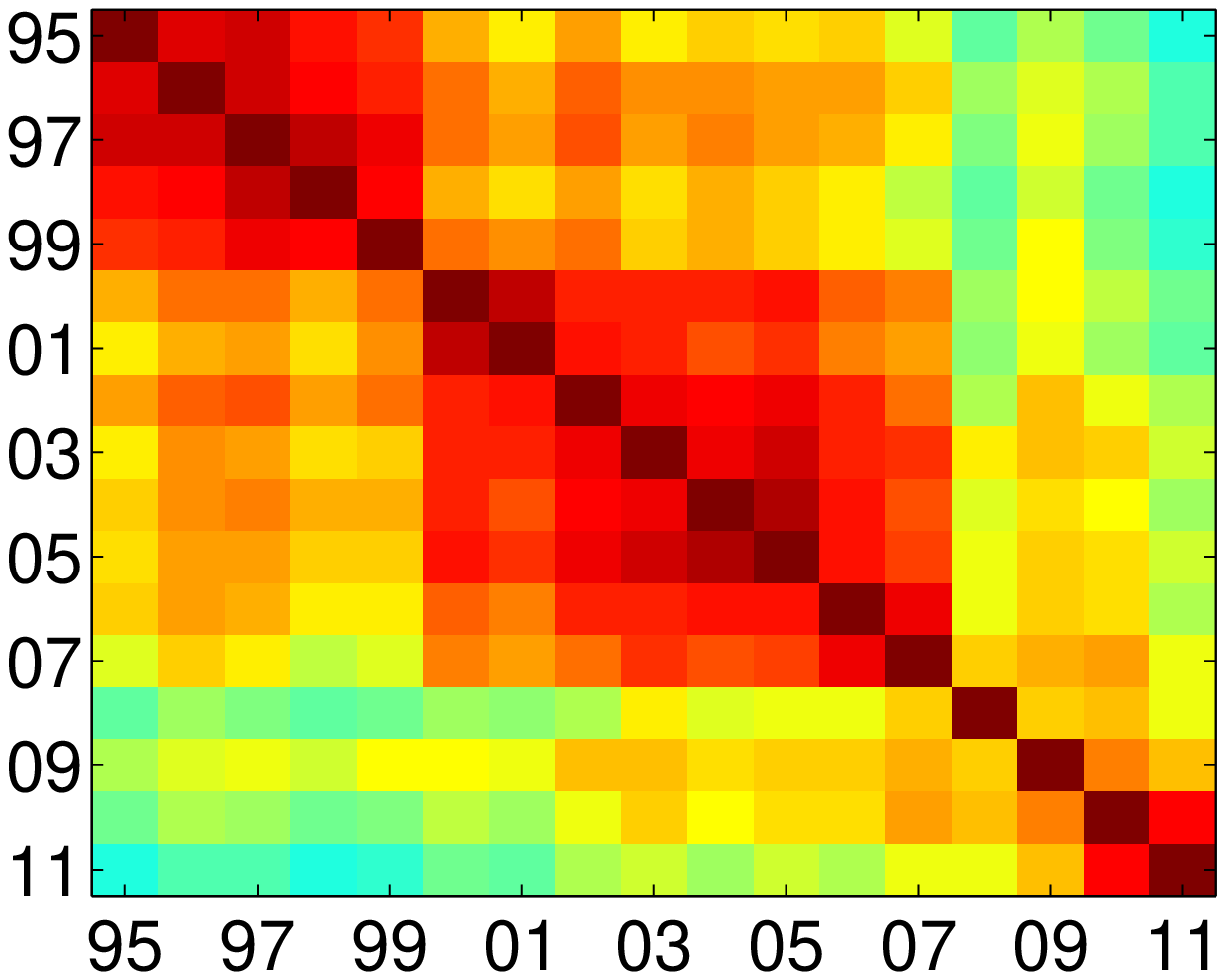}
\caption{USA}
\label{fig:USA}
\end{subfigure}
\begin{subfigure}{0.34\textwidth}
\centering
\includegraphics[width=0.7\textwidth]{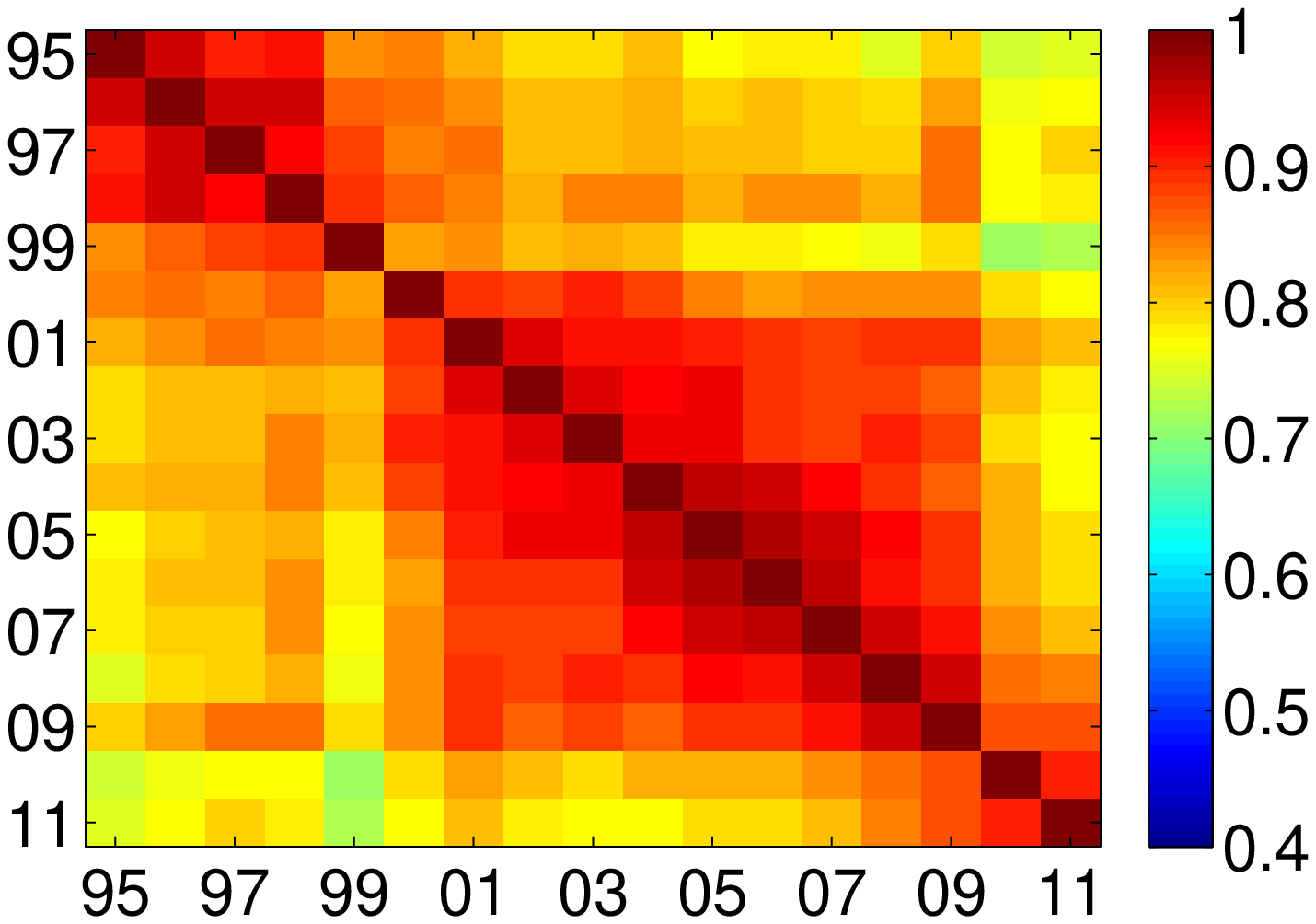}
\caption{Germany}
\label{fig:Germany}
\end{subfigure}
\caption{Kendall correlation coefficient heat map shows the industry rank
  correlation coefficient between each pair of years from
  $1995-2011$. Using Kendall correlation coefficient $\tau$,
  we investigate the evolution of
  the economic structure of the investigated countries. For each year,
  we rank the industries in each country according to their tolerance
  values, and then calculate the Kendall $\tau$ for every pairs of years. We
  plot the values of the Kendall $\tau$ for all pairs of years, for China
  (top), USA (middle), and Germany (bottom), using a color code ranging
  from blue, for low similarity, to red, for high similarity (see
  Supplementary Information for all other countries).}
  \label{3plot_kendall}
\end{figure*}

We use the Kendall $\tau$ to investigate the evolution of the economic
structure of each country in our sample. For each year, we rank the
industries in each country according to their tolerance values, and
calculate the Kendall $\tau$ for every year
pair. Figure~4 shows the values of the Kendall $\tau$
for all pairs of years, for China (Fig.~4(a)), USA
(Fig.~4(b)), and Germany (Fig.4(c)),
using a color code ranging from blue, for low similarity, to red, for
high similarity (see Supplementary Information for all other countries).
For these three countries, we find different behaviors in terms of the
stability and consistency of the economic structures.
For the case of China, we observe that the structure changes significantly,
with high values of the Kendal correlation (represented using red)
only presenting for previous 2$\sim$3 years (as can be observed from the
diagonal of elements of Fig.~4(a)). However,
in the case of the USA, it is possible to observe three periods in terms of stability of the economic structure: 1995-1999, 2000-2007, and 2008-2011 (see Fig.~4(b)). The first marks the period leading to the ''dot.com'' crisis, which was followed by a significant change in the US market structure. The second marks the period leading unto the financial crisis, which again was followed by a significant change in the US market structure. Finally, the third marks the period following the financial crisis, which shows no stable period in terms of market structure. Finally, in the case of Germany, we observe only two periods: 1995-1999, and 2000-2011 (see Fig.~4(c)). The first period can be also attributed to that apparent in the case of the USA, however possibly to a lesser extent and also resulting from the introduction of the Euro currency. However, in the case of Germany we do not observe a change in structure following the recent financial crisis, which highlights the degree of stability in the German economy.

\section{The rise of China}

Due to the fact that economic influence is dynamic across time, we ask whether the
methodology presented here can provide new information on the increase
or decline of economic importance. In each year
we calculate the individual industry tolerance as described above. We then
calculate the average tolerance of each country for a given year.
Figure~5 shows the average of $p_c$ of country for
the 17-year period investigated. In Figure~5(a), we show the largest tolerance $p_c$ in China, the USA and
  Germany along 17 years. Figure~5(b) shows the average of 4 largest industries $p_c$, and Figure~5(c) shows the average of 8 largest $p_c$ in each country.

  We find that the average tolerance $p_c$ of China becomes larger than that of the USA
  after 2003, which is most pronounced in Figure~5(b). The USA tolerance $p_c$ first increases from 1995 to 2000, then
  decreases from 2000 to 2009, with a slight increase in 2010-2011. Germany's tolerance $p_c$ in general increase in the investigated period and shows certain fluctuations
  between 2000 and 2005. Note that for the USA the change across
time is minor but that the economic importance of China increases
significantly. The economic importance of China relative to that of the
USA shows a consistent increase from year to year, illustrating how the
economic power structure in the world's economy has been changing during time.

\begin{figure}
\centering
\includegraphics[width=0.6\textwidth]{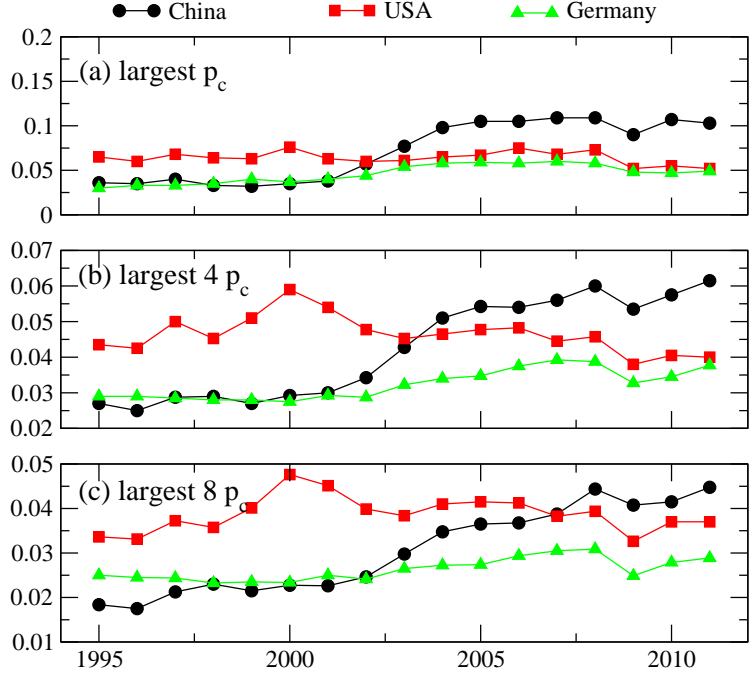}
\caption{Tolerance $p_c$ changes of China, the USA and Germany for 17
  years. We show that the average tolerance $p_c$ increases from 1995 to
  2011. In (a), we shows the largest tolerance $p_c$ in China, the USA and
  Germany along 17 years. (b) Shows the average of 4 largest $p_c$ in each country. We can see the $p_c$ of China becomes larger than the USA
  after 2003.
  (c) Shows the average of 8 largest $p_c$ in each country. The difference
  between the average $p_c$ of China and that of the USA is smaller compared to
  that in (b) in years 2005-2011. This is because the average includes more industries
  with small $p_c$ and mainly the importance of large industries increase.
  The USA tolerance $p_c$ first increases from 1995 to 2000, then
  slightly decreases from 2000 to 2009, with a small increase in 2010-2011.
  Germany's tolerance $p_c$ in
  general slightly increases in the 17-year period and shows small fluctuations
  between 2000 and 2005. }
\label{17yearPc3ctry.eps}
\end{figure}

To further validate these results, we compare the total product output (see Figure~6, red triangle) and average tolerance $p_c$ (see Figure~6, black circles) for China, USA, and Germany, as a function of time. The product output (see Figure~6, red triangle) value is the total money flow a country supplies
to the other countries plus value added in the products, which also indicates its total trade impact on foreign countries. Studying Figure~6, we find that generally speaking the total outputs of the three investigated countries grows throughout time, with that of the USA and China being higher in value than that of Germany (see Figure~6, right y-axis). However, by comparing the tolerance $p_c$, we find three different behaviors. First, for the case of China (Figure~6(a)), we find that both the tolerance $p_c$ and total outputs are growing in time, and observe that in the early 2000's there was a jump in the tolerance $p_c$, following by a sharp increase in the total output. Secondly, for the case of the USA (Figure~6(b)), we find that while the total output is increasing in time, the tolerance $p_c$ is in a decreasing trend. Finally, for the case of Germany (Figure~6(c)), we find that while the total import is increasing in time, the tolerance $p_c$ is rather stable with small fluctuations.

\begin{figure}[h]
\centering
\includegraphics[width=0.6\textwidth]{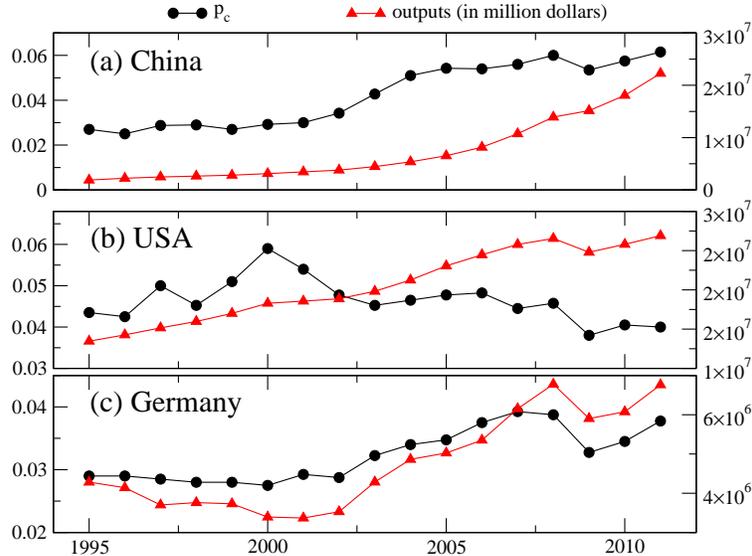}
\caption{Tolerance $p_c$ of China, the USA and Germany comparing to
the total product output value. For each country, the $p_c$ is an
average of the largest four industry $p_c$ of this country (black circles).
The product output (red triangle) value is the money flow a country supplies
to the rest of the countries, which also indicates its impact to foreign countries.
In (a) and (c), the trends of the $p_c$ and output value are similar, which indicate that
China and Germany become more influential to the world economy. In (b),
the $p_c$ of the USA increases in the period of year 1995 to year 2000, and
decreases from 2000 to 2011, while the output value increases from 1995 to
2011 in general.}
\label{3ctrymoney17}
\end{figure}

This provides further evidence into the change in influence of these three important economies - that of China is increasing, that of the US is declining, and that of Germany is rather constant across time. Comparing the total output to the tolerance $p_c$ provides further evidence that the tolerance measurement of country's impact reveals
the underlying economic evolving dependencies which is not obvious from the simple measurement
of total output capability.

\section{Conclusions}
We developed a framework to quantify
interdependencies in the world industrial network and
measure risk levels in global markets. We use the
methodology to rank the economic importance of
each industry and country according to the global damage that would result from
their failures. Using network science to investigate input-output data of money flows between different economic industries, it becomes possible to stress test the global economic network and identify vulnerabilities and sources of systemic risk. Our quantitative results shed new light on China's
increasing economic influence over other economies, including that of
the USA. The resulting information will enable policy and decision
makers to better measure, understand, and maintain financial
stability.

\begin{acknowledgments}
We would like to thank the European Commission FET Open Project
FOC 255987 and FOC-INCO 297149 for financial support.
We wish to thank the ONR, DTRA, DFG, EU project,
Multiplex and LINC, EU projects, the Keck Foundation, and the Israel Science Foundation
for financial support.
\end{acknowledgments}

\end{document}